\documentclass{grf2002}
\usepackage{graphicx,amssymb}
\journal{Gravity Research Foundation 2002}
\begin{document}
\addtolength{\topmargin}{+50pt}

\def\beq{\begin{eqnarray}}
\def\eeq{\end{eqnarray}}

\def\g{\mathbf g}
\def\A{{\mathcal A}}

\def\B{{\mathbf B}}
\def\E{{\mathbf E}}
\def\SS{{\mathcal S}}
\def\X{{\mathcal X}}

\def\F{{\mathcal F}}
\def\R{{\mathcal R}}
\def\W{{\mathcal W}_\mu}

\def\s{\mbox{\boldmath$\displaystyle\mathbf{\sigma}$}}
\def\et{\mbox{\boldmath$\displaystyle\mathbf{\eta}$}}
\def\S{\mbox{\boldmath$\displaystyle\mathbf{S}$}}

\def\J{\mbox{\boldmath$\displaystyle\mathbf{J}$}}
\def\K{\mbox{\boldmath$\displaystyle\mathbf{K}$}}
\def\P{\mbox{\boldmath$\displaystyle\mathbf{P}$}}
\def\p{\mbox{\boldmath$\displaystyle\mathbf{p}$}}
\def\hp{\mbox{\boldmath$\displaystyle\mathbf{\widehat{\p}}$}}
\def\x{\mbox{\boldmath$\displaystyle\mathbf{x}$}}
\def\0{\mbox{\boldmath$\displaystyle\mathbf{0}$}}
\def\bv{\mbox{\boldmath$\displaystyle\mathbf{\varphi}$}}
\def\hbv{\mbox{\boldmath$\displaystyle\mathbf{\widehat\varphi}$}}
\def\bn{\mbox{\boldmath$\displaystyle\mathbf{\nabla}$}}
\def\bl{\mbox{\boldmath$\displaystyle\mathbf{\lambda}$}}
\def\bl{\mbox{\boldmath$\displaystyle\mathbf{\lambda}$}}
\def\br{\mbox{\boldmath$\displaystyle\mathbf{\rho}$}}
\def\1{1}
\def\bfhh{\mbox{\boldmath$\displaystyle\mathbf{(1/2,0)\oplus(0,1/2)}\,\,$}}
\def\mn{\mbox{\boldmath$\displaystyle\mathbf{\nu}$}}
\def\amn{\mbox{\boldmath$\displaystyle\mathbf{\overline{\nu}}$}}
\def\mne{\mbox{\boldmath$\displaystyle\mathbf{\nu_e}$}}
\def\amne{\mbox{\boldmath$\displaystyle\mathbf{\overline{\nu}_e}$}}
\def\rlh{\mbox{\boldmath$\displaystyle\mathbf{\rightleftharpoons}$}}

\def\wm{\mbox{\boldmath$\displaystyle\mathbf{W^-}$}}
\def\bhh{\mbox{\boldmath$\displaystyle\mathbf{(1/2,1/2)}$}}
\def\hh{$(1/2,1/2)\;$}
\def\h00h{\mbox{\boldmath$\displaystyle\mathbf{(1/2,0)\oplus(0,1/2)}$}}
\def\znbb{\mbox{\boldmath$\displaystyle\mathbf{0\nu \beta\beta}$}}


\begin{frontmatter}


\title{On the spin of gravitational bosons}


\author{D.\ V.\ Ahluwalia}$^a$,
\ead{ahluwalia@heritage.reduaz.mx}
\author{N.\ Dadhich}$^b$,
\ead{nkd@iucaa.ernet.in}
\author{M.\ Kirchbach}$^a$,
\ead{kirchbach@chiral.reduaz.mx}
\address{$^a$ Theoretical Physics Group, 
Fac. de Fisica de la UAZ, Zacatecas, Ap. Postal C-600,
ZAC 98062, Mexico.}
\address{$^b$ Inter-University Center for Astronomy and Astrophysics (IUCAA),
Post Bag 4, Ganeshkhind
Pune 411 007, India.}

\begin{abstract}
We unearth spacetime structure of massive vector bosons, gravitinos, and
gravitons. While the curvatures associated with these particles
carry a definite spin, the underlying potentials cannot be, and should
not be, interpreted as single spin objects. For instance,
we predict that a spin measurement in the rest frame of a massive gravitino
will yield the result $3/2$ with probability one half, and $1/2$ with
probability one half.
The simplest scenario leaves the Riemannian curvature 
unaltered; thus avoiding conflicts with classical tests of 
the theory of general relativity. However, the quantum structure acquires
additional contributions to the propagators, and it
gives rise to additional phases. 
\end{abstract}

\end{frontmatter}

\newpage

{\sc I. Interface of quantum and gravitational realms}

The conceptual foundations of the general theory of relativity were 
established in an era when the quantum revolution had yet to fully 
inseminate the thinking of those desiring a unification of then-known 
interactions, or those working on a  quantum theory of gravity. In that 
early part of the last  century a crucial lesson from the unification 
of the electroweak theory was still several decades in 
the future. This circumstance has now given rise to a situation where
a brute force quantization of gravity has exhausted the efforts of
its pioneers, and may be considered to have failed. 
In the vacuum left by this failure two fundamentally new notions have
arisen. These are the introduction of {\em extended objects\/} (strings, etc.)
and {\em supersymmetry.\/} The former asks for abandoning the concept
of point particles, while the latter is a natural extension of spacetime
symmetries that places fermions and bosons at the same formal footing.

In recent years
it has been realized that if one incorporates classical general-relativistic
framework in certain {\em Gedanken\/} quantum measurement processes \cite{as} 
then the wave-particle duality is modified \cite{grf1994,akempf,pla2000}. 
For a one-dimensional motion, such a modification  
may  be encoded in an expression of the form:
\beq
\lambda=\frac{\overline \lambda_P}{\tan^{-1}\left(\overline{\lambda}_P/
\lambda_{dB}\right)}
\cases{\rightarrow\lambda_{dB}&\quad\mbox{for low energy regime} \cr
\rightarrow 4 \lambda_P&\quad\mbox{for Planck regime} }\,,
\label{mwpd}
\eeq
where
\beq
\lambda_P=\sqrt{\frac{\hbar G}{c^3}}\,,\quad\lambda_{dB}=\frac{h}{p}\,,
\eeq
are in turn the Planck length and the de Broglie wavelength; while
$\overline{\lambda}_P=2\pi \lambda_P$ is the Planck circumference.
 
The same set of {\em Gedanken\/} experiments implies a non-commutative
nature of spacetime which has been extensively studied in a series of papers
\cite{jm,sh,gac,mm,ns,fs}.   The modification of the wave-particle
duality at the Planck scale renders the notion of a point particle
operationally meaningless. Therefore,
the latter must be replaced by some,
yet undefined, {\em fuzzy spacetime entity,\/} 
which consistently captures in it the 
fundamental non-commutative nature of spacetime, and  the modified 
wave-particle  duality [of which Eq. (\ref{mwpd}) is only an approximate 
one-dimensional representation]. We  shall argue in the concluding
remarks  that Eq. (\ref{mwpd}) can be studied in terrestrial laboratories.
At the same time a recent Amelino-Camelia proposal  allows
to probe the associated spacetime fuzziness \cite{gacn,dvan}.  

In order for the proposed fuzzy spacetime entities to possess
well-defined macroscopic properties they must transform 
from one inertial frame to another via various representations
of the Lorentz group. The fuzzy nature {\em cannot\/} be captured
by form factors so useful in describing QCD's extended objects such as
protons, neutrons, and hadrons in general. 
The protons and the neutrons
are described  by Dirac's $(1/2,0)\oplus(0,1/2)$ representation space,
and the remaining hadrons -- irrespective of detailed
QCD considerations -- must also transform according one, or the other,
representations of the Lorentz group. In addition, 
QCD solutions associated with these
particles require relativistically covariant form factors
that encode in them the extended 
nature of these objects. However, for the 
fuzzy spacetime entities the extendedness is characterized by the
Planck length, $\lambda_P$, and  it cannot be probed in the same manner
as, e.g., a nucleon charge distribution. 
It is prevented by the general-relativistically
modified wave-particle duality which saturates the matter wavelengths to
$\lambda_P$ (or, perhaps a few times $\lambda_P$).

At this stage a critical reader may ask: Are there {\em any\/} hints
for the existence of fuzzy spacetime entities, and for supersymmetry? 
A purely formal answer to that question is: No. 
Nonetheless, elements of both spacetime fuzziness
and supersymmetry seem to be present in particle physics:

\begin{enumerate}
\item
Spacetime
fuzziness is hinted ever  since the discovery
of CP violation in the neutral kaon system.
Also recent set of strong indications
for flavor oscillations in the neutrino sector
provide an independent support of this observation.
Indeed, oscillation phenomena indicate towards the fact that -- 
for reasons not yet fully understood -- kaons and
neutrinos are not mass eigenstates (to be identified with eigenstates
of the first Casimir invariant of the Poincar\'e spacetime group). 
Their masses carry an inherent fundamental
uncertainty, thus providing an example of a well-established 
fuzziness at the spacetime level that is consistent with 
principles of quantum framework.

\item
On the other side,
in its simplest form, the algebra of supersymmetry contains besides the
ten Poincar\'e group generators, also
four anticommuting generators, the  ``supertranslations,''
which are components of a Majorana spinor. So, it comes about that spacetime
structure of supersymmetry finds itself deeply intertwined with the Majorana
aspect. In this context, recent results of Klapdor-Kleingrothaus 
{\em et al.\/} \cite{hk,ew,ak}, which provide a first direct evidence for 
neutrinoless double  beta decay, $0\nu\beta\beta$, are of particular 
interest. In its simplest interpretation, the $0\nu\beta\beta$
experimental signal arises from the Majorana nature of $\nu_e$. Even though 
the experiment by itself does not necessarily require a supersymmetric
framework for its explanation, the indication towards a Majorana spacetime 
structure suddenly acquires a pivotal importance as a structure at the heart
of supersymmetry. 
\end{enumerate}

While fuzziness of spacetime has important and widely discussed
consequences for a quantum theory of gravity, and while this
aspect is in fact a natural consequence of the interplay of the
gravitational and quantum realms, it is our opinion that 
any successful framework for a quantum theory of gravity must
incorporate a further lesson, namely the one taught by the 
unification of the electroweak theory. 
It is that very lesson on which the present essay is focused. 

To briefly outline the task, recall that
the massive gauge bosons of the electroweak theory 
-- as regards their spacetime structure --
have to transform according to the  $(1/2,1/2)$ representation space,
\beq
\A^\mu(x):\quad(1/2,1/2)\,.
\eeq
The latter, in being a four dimensional representation space,
requires four independent degrees of freedom.
Despite that demand, spacetime structure of
massive gauge fields is treated within
Proca's framework  captured by the  equation,
\beq
\partial_\mu \F^{\mu\nu}(x) + m^2 \A^\nu (x) = 0\,.\label{peq}
\eeq
Due to the antisymmetric nature of the Proca curvature (or,
more commonly called field strength tensor) $\F^{\mu\nu}(x)$, one in fact 
restricts to only three degrees of freedom. This happens
because on taking the divergence of the above equation, and exploiting the
stated antisymmetric nature of $\F^{\mu\nu}(x)$, the solutions are 
restricted to a subset of divergence-free
$(1/2,1/2)$ space solutions, i.e. to $\A^\nu(x)$ that satisfy
\beq
\partial_\nu \A^\nu(x)=0\,.\label{dmuamu}
\eeq  
In this way, being mainly guided  by the belief that the 
$(1/2,1/2)$ representations space,
despite its obvious four-dimensional character, ought to carry only
three physical degrees of freedom -- Proca's framework attempts to force a  
single spin-one interpretation upon the $(1/2,1/2)$ representation space.  

On the other side, the renormalizability of the electroweak theory required 
to supplement the Proca propagator by an additional -- at that time
ad hoc -- St\"uckelberg term -- which  was thought to arise 
from an unusual  scalar field of a negative norm
outside the $(1/2,1/2)$ space under consideration. In addition,
the propagator of this scalar had to be brought in with a ``wrong''
sign \cite{Veltman} relative to Proca's propagator
in order to guarantee vanishing of divergences in the theory. 
This is well recounted by Veltman 
in his recent Nobel Prize lecture \cite{Veltman}.
Unexpectedly, while its necessity is fully realized,
the physical and mathematical origin of the St\"uckelberg term has not been
understood at its fundamental representation-theory level.

So the purpose of this essay now is to trace back the origin of the 
St\"uckelberg term to the completeness of the $(1/2,1/2)$ representation 
space and to show how the electroweak gauge bosons
come to carry an indefinite spin. Once we go through that argument
it will be apparent to the reader 
how to move on to gravitinos and gravitons;
and   how they too cannot carry a definite spin as long as one allows 
them to be endowed with a non-zero mass (however small).\footnote{ Already,
graviton masses and lower-spin components in gravity
are of interest in astrophysical and cosmological
contexts; for instance, due to indications for an accelerating 
universe \cite{acc,tm,ds,nd1,nd2}.
}
Note that
the gauge bosons in a  supersymmetric 
quantum theory of gravity are gravitinos and gravitons.
From an observational point of view the data on pulsars PSR B1913+16 and
PSR B1534+12 requires graviton mass to be less than $7.6\times 10^{-20}
\,\,\mbox{eV}$ at $90\%$ confidence \cite{fs2}, while the gravitino mass
may be tens of orders of magnitude higher. 
The gravitinos and gravitons transform in turn as:
\beq
&&\psi^\mu(x):\quad\Big[(1/2,0)\oplus(0,1/2)\Big]\otimes(1/2,1/2)\,,\\
&&g^{\mu\nu}(x): \quad (1/2,1/2)\otimes(1/2,1/2)\,.
\eeq

Before proceeding further, we wish to draw readers
attention to a point that appears crucial to further 
understanding. We wish to emphasize that all
the peculiarities of the vector potentials mentioned above
carry  relevance solely at the quantum level.
Recall the classical work by Aharonov and Bohm \cite{ab} 
(for the massless limit), showing that the potentials 
give rise to  non-trivial and observable quantum phases which
do not have any classical counterpart. 
Moreover, the indefinite spin of the vector potentials  
so important for bringing into the propagator the terms
needed for renormalizability, contrasts 
the case of $\F^{\mu\nu}(x)$, that is endowed by a unique spin, i.e., with
spin one.
As long as $\F^{\mu\nu}(x)$ encodes the forces
acting on test particles, the multi-spin character
of the vector potentials will leave the classical level of the theory
unaltered.

{\sc II. Mathematical structure of the \hh representation space}

Spinors and Lorentz vectors  play a pivotal role
in physics. The latter embody in them
the transformation properties of the \hh representation space
which, by definition,  is the direct product of
the $(1/2,0)$ and $(0,1/2)$ representation spaces:
\beq
(1/2,1/2): \quad(1/2,0)\otimes(0,1/2)\,.
\eeq

We shall work in the momentum space.

In the rest frame, where the
three-momentum of the particle under
consideration is zero, $\p=\0$; 
the \hh representation space decomposes into two
subspaces of spin one, and  spin zero. 
These subspaces are spanned by four vectors:
\beq
&& \A_a^{1,+1}(\0) = h^+\otimes h^+\,,\label{eq1}\\
&& \A_a^{1,0}(\0)  = \frac{1}{\sqrt{2}}\left(h^+\otimes h^- + h^-\otimes
h^+\right)\,,\label{eq2}\\
&& \A_a^{1,-1}(\0) = h^-\otimes h^-\,,\label{eq3}\\
&& \A_a^{0,0}(\0)  = \frac{1}{\sqrt{2}}\left(h^+\otimes h^- - h^-\otimes
h^+\right)\,.\label{eq4}
\eeq
In Eqs.  (\ref{eq1})-(\ref{eq4}), $h^\pm$ are eigenstates of the
spin-$1/2$ helicity operator, $(\s/2)\cdot\hp$
\beq
h^+=m^{1/4} 
\left( 
\begin{array}{c}
\cos(\theta/2) e^{-i\phi/2}\\ 
\sin(\theta/2) e^{i\phi/2}
\end{array}
\right)\,,\quad
h^-=  m^{1/4} 
\left( 
\begin{array}{c}
\sin(\theta/2) e^{-i\phi/2}\\ 
- \cos(\theta/2) e^{i\phi/2}
\end{array}
\right)\,.\label{eq6}
\eeq
Here, $\s=(\sigma_x,\sigma_y,\sigma_z)$; $\sigma_i$ are the 
usual Pauli matrices, and $\hp$ is the  unit momentum vector 
with Cartesian components 
	$(\sin(\theta) \cos(\phi),\;
         \sin(\theta)\sin(\phi),\;\cos(\theta))$.
The factor $m^{1/4}$, besides other matters,  
allows, in the massless limit, the  $\A_a^{s,h}(\0) $ to identically vanish. 
Since massless particles have no rest frame, the preceding 
is a physical requirement.
The subscript, $a$, is a Lorentz index (however, see Eq. (\ref{s}) below).
The superscripts  are defined as:
\beq
\S^2 \; \A_a^{s,h}(\0) = s(s+1)\; \A_a^{s,h}(\0)\,,\quad
\S\cdot\hp \;\A_a^{s,h}(\0) = h  \;\A_a^{s,h}(\0)\,.\label{eq5}
\eeq

The  generators of rotation for the \hh representation space [that appear 
in Eq. (\ref{eq5})] are:
\beq
S_x=\frac{1}{2}
\left(
\begin{array}{cccc}
0 & 1 & 1 & 0 \\
1 & 0 & 0 & 1\\
1 & 0 & 0 & 1\\
0 & 1 & 1 & 0 \\
\end{array}
\right)\,,\;
S_y=\frac{1}{2}
\left(
\begin{array}{cccc}
0 & -i & -i & 0 \\
i & 0 & 0 & -i\\
i & 0 & 0 & -i\\
0 & i & i & 0 \\
\end{array}
\right)\,,\;
S_z=
\left(
\begin{array}{cccc}
1 & 0 & 0 & 0 \\
0 & 0 & 0 & 0\\
0 & 0 & 0 & 0\\
0 & 0 & 0 & -1 \\
\end{array}
\right)\,.
\eeq
The application of the boost\footnote{The $(1/2,0)$- and $(0,1/2)$-
boosts that appear in the equation below are:
\[
\kappa^{\left(\frac{1}{2},0\right)}=\exp\left(+\, 
\frac{\s}{2}\cdot\bv\right)= \sqrt{\frac{E+m}{2\,m}}
\left(\1_2+\frac{\s\cdot\p}{E+m}\right)
\,,
\]
\[
\kappa^{\left(0,\frac{1}{2}\right)}=\exp\left(-\, 
\frac{\s}{2}\cdot\bv\right)= \sqrt{\frac{E+m}{2\,m}}
\left(\1_2-\frac{\s\cdot\p}{E+m}\right)\,.
\]
The boost parameter, $\bv$, is defined as:
\[
\cosh(\varphi)=\frac{E}{m},\quad
\sinh(\varphi)=\frac{\vert\p\vert}{m},\quad 
{\hbv}
=\frac{\p}{\vert \p\vert}\,.
\]
We use the notation in which 
$\1_n$ and $0_n$ represent
$n\times n$ identity and null matrices, respectively. 
}
\beq
\kappa^{\left(\frac{1}{2},\frac{1}{2}\right)} = 
\kappa^{\left(\frac{1}{2},0\right)}\otimes
\kappa^{\left(0,\frac{1}{2}\right)},
\eeq
to the  $\A_a^{s,h}(\0) $ yields, in the order presented in Eqs. 
(\ref{eq1})-(\ref{eq4}),
\beq
&& A_a(\p,\,1)
= \frac{\sqrt{m}}{2}
\left(
\begin{array}{c}
2 e^{-i \phi} \cos^2(\theta/2) \\
\sin(\theta) \\
\sin(\theta)  \\
2 e^{i \phi} \sin^2(\theta/2)
\end{array}
\right)\,,\;
 \A_a(\p\,,2)
= \frac{1}{\sqrt{2\,m}} 
\left(
\begin{array}{c}
 e^{-i \phi} E \sin(\theta) \\
- \left(\vert \p \vert + E \cos(\theta)\right) \\
\vert \p \vert - E \cos(\theta) \\
- e^{i \phi} E \sin(\theta)
\end{array}
\right)\,,\nonumber\\
&& \A_a(\p\,,3)
=  \frac{\sqrt{m}}{2} 
\left(
\begin{array}{c}
2 e^{-i \phi} \sin^2(\theta/2) \\
- \sin(\theta) \\
- \sin(\theta)  \\
2 e^{i \phi} \cos^2(\theta/2)
\end{array}
\right)\,,\;
 \A_a(\p,\,4)
= \frac{1}{\sqrt{2\,m}} 
\left(
\begin{array}{c}
 e^{-i \phi} \vert \p\vert  \sin(\theta) \\
- \left( E + \vert \p \vert \cos(\theta)\right) \\
E - \vert \p \vert  \cos(\theta) \\
- e^{i \phi} \vert \p\vert  \sin(\theta)
\end{array}
\right)\,.\nonumber\\\label{xi}
\eeq
The reason for change in notation from $\A^{s,h}(\0)$ to $\A(\p,\,\zeta)$,
$\zeta=1,2,3,4$, shall be made apparent in Sec. III.1 below.

In a parallel to the Dirac's $(1/2,0)\oplus(0,1/2)$ representation space, 
and to emphasize  certain non-trivial mathematical similarities, 
we introduce 
\beq
\lambda^{00}
=\left(
\begin{array}{cccc}
-1 & 0  & 0 & 0 \\
0  & 0  & -1 & 0\\
0  & -1 &  0 & 0\\
0  & 0  & 0 & -1
\end{array}
\right),
\eeq
and define:
\beq
\overline{\A}_a(\p)= \left(\A_a\right)^\dagger\lambda^{00}\,.
\eeq
Then, the orthonormality and completeness relations
for the \hh  rep\-resenta\-tion space read (with no summation intended
on the Lorentz index):

\beq
&&\overline{\A}_a (\p\,,\zeta) \;
  \A_a (\p\,,{\zeta^\prime}) 
=\cases{ -  \,m\,\delta_{\zeta\zeta^\prime}, 
\;\;\mbox{for}\;\zeta=1, 2, 3
\nonumber\cr
+ \,m\,\delta_{\zeta\zeta^\prime},
\;\;\mbox{for}\;  \zeta=4}\,,\\
&&\frac{1}{m}\left[ \A_a(\p,\,4)\;\overline{\A}_a(\p\,,4) -
\sum_{\zeta=1,2,3}  
\A_a(\p,\,{\zeta})\;\overline{\A}_a(\p\,,{\zeta^\prime})
\right]=\1_4\,. 
\eeq

The Lorentz index, $a$, that appears in the above expressions
is not the usual ``time, space'' (i.e., usual 0,1,2,3) index.
The latter, denoted by $\mu,\nu\ldots$, 
is obtained via the following transformation,
\beq
 \A^\mu(\p) = \SS^{\mu a} \A_a (\p) \,, \label{s}
\eeq
with 
\beq
\SS=\frac{1}{\sqrt{2}}
\left(
\begin{array}{rrrrrrr}
0 & &i & &-i && 0 \\
-i & &0 && 0 && i\\
1 & &0 && 0 & &1\\
0 & &i && i & &0
\end{array}
\right)\,.
\eeq
These satisfy a new wave equation \cite{ka}:
\beq
\left(\Lambda_{\mu\nu} 
 p^\mu p ^\nu \pm m^2 I_4\right) \A(\p\,,\zeta)=0\,,\label{Eq} 
\eeq
where the plus sign is to be taken for
$\zeta=1,2,3$, while the minus sign is for $\zeta=4$.

The $\Lambda_{\mu\nu}$ 
matrices are: $\Lambda_{00}=\mbox{diag}(1,-1,-1,-1)$,
$\Lambda_{11}=\mbox{diag}(1,-1,1,1)$, $\Lambda_{22}=\mbox{diag}(1,1,-1,1)$,
$\Lambda_{33}=\mbox{diag}(1,1,1,-1)$, and 
\beq
&& \Lambda_{01}=
\left(\begin{array}{cccc}
0 & -1 & 0 & 0 \\
1 & 0 & 0 & 0 \\
0 & 0 & 0 & 0\\
0 & 0 & 0 & 0
\end{array}\right),\,\,
\Lambda_{02}=
\left(\begin{array}{cccc}
0 & 0 & -1 & 0 \\
0 & 0 & 0 & 0 \\
1 & 0 & 0 & 0\\
0 & 0 & 0 & 0
\end{array}\right), \,\,
\Lambda_{03}=
\left(\begin{array}{cccc}
0 & 0 & 0 & -1 \\
0 & 0 & 0 & 0 \\
0 & 0 & 0 & 0\\
1 & 0 & 0 & 0
\end{array}\right), \,\,\nonumber\\
&& \Lambda_{12}=
\left(\begin{array}{cccc}
0 & 0 & 0 & 0 \\
0 & 0 & -1 & 0 \\
0 & -1 & 0 & 0\\
0 & 0 & 0 & 0
\end{array}\right), \,\,
\Lambda_{13}=
\left(\begin{array}{cccc}
0 & 0 & 0 & 0 \\
0 & 0 & 0 & -1 \\
0 & 0 & 0 & 0\\
0 & -1 & 0 & 0
\end{array}\right),\,\,
\Lambda_{23}=
\left(\begin{array}{cccc}
0 & 0 & 0 & 0 \\
0 & 0 & 0 & 0 \\
0 & 0 & 0 & -1\\
0 & 0 & -1 & 0
\end{array}\right)\,.
\eeq
They are symmetric in the Lorentz index, $ \Lambda_{\mu\nu} = 
\Lambda_{\nu\mu}$.\footnote{Parenthetically, we note that the 
$\SS$-transformed
$\lambda^{00}$, i.e., $\SS \lambda^{00} \SS^{-1}$, 
equals $\Lambda_{00}$ and is nothing but the standard
flat-spacetime metric, $\g$, with the signature $(1,-1,-1,-1)$.}
The  interchange,
$\A^\mu(\p,\,\zeta=1,2,3) \rightleftharpoons  
\A^\mu(\p,\,\zeta=4)$, corresponds to  
\beq
m\rightleftharpoons
i\, m\,,
\eeq
This circumstance  suggests that the ``negative mass squared'' term 
in spontaneous symmetry breaking may have the above spacetime 
structure at its origin.

{\sc III. Physical structure of the \hh representation space}

{\bf 1.} The  $\A(\p,\,\zeta)$,  are, in general, not eigenstates
of $\S^2$. This is because the 
$\kappa^{\left(\frac{1}{2},\frac{1}{2}\right)}$ does {\em not,\/} 
for an arbitrary $\A(\p\,)$,  commute with
$\S^2$. 

If spin is to be identified with eigenvalues of $\S^2$, then 
$(1/2,1/2)$ representation space
does not carry a single-spin interpretation. Spin-1 and Spin-0 
thus become covariantly inseparable in the  
$(1/2,1/2)$ representation space. On the other hand, if one
wishes to have a {\em pure\/} spin-1 massive object, 
then that object must transform according to the
$(1,0)\oplus(0,1)$ representation \cite{bww}. 

Even though, in their rest frame $\A(\p,\,\zeta=1,2,3)$ 
and $\A(\p,\zeta=4)$ appear as
spin one and spin zero objects, respectively;  they 
cannot be identified with  spin one  and ``scalar (i.e. spin $0$)'' particles
in an arbitrary frame.

{\bf 2.} The  $\A(\p,\,\zeta=1,2,3)$,
coincide with the solutions of Proca framework (and are divergence-free); 
whereas   $\A(\p,\,\zeta=4)$, 
is divergence-full:
\beq
&& p_\mu \A^\mu(\p,\,\zeta=1,2,3) = 0\,,\\
&&  p_\mu \A^\mu(\p,\,\zeta=4)= i m^{3/2}\,.
\eeq
{\em Thus mass serves as a source of  $\A^\mu(\p,\,\zeta=4)$.}  

{\bf 3.} On quantization, we find that the
{\em numerator\/} of the propagator associated with the 
$\zeta=1,2,3$ sector (Proca sector) is:
\beq
\frac{1}{m}\sum_{\zeta=1,2,3}
\SS\left(\A^\mu(\p,\zeta)\; \overline\A^\nu(\p,\,\zeta) \lambda^{00}\right)
\SS^{-1} = -\,g^{\mu\nu} + \frac{p^\mu p^\nu}{m^2}
\eeq
while the contribution to the numerator of the propagator from the 
$\zeta=4$ sector (St\"uckelberg sector) turns out to be: 
\beq
\frac{1}{m}\SS\left(\A^\mu(\p,\,4)\; {\overline\A}^\nu(\p,\,4) \lambda^{00}\right)
\SS^{-1} = \quad\frac{p^\mu p^\nu}{m^2}
\eeq
The latter contribution lies outside the Proca framework,
and is the {\em key\/} ingredient in the renormalizability of 
the electroweak unification. Here it appears naturally --
with the same physical interpretation that mass is 
its source. Here, as well as in the electroweak unification,
the relative sign of the two contributions is opposite to
that which naive vacuum expectation value of the relevant time-ordered
field operators implies.


{\bf 4.} The $\A^\mu(\p,\,\zeta=1,3)$,
 correspond 
to left- and right- circular polarizations, 
while  the  $\A^\mu(\p,\,\zeta=2,4)$ 
are the longitudinal, and time-like polarizations, respectively.

In order to construct the field strength tensor, $\F^{\mu\nu}(\p)$, 
we provide the relation between, $\A^\mu(\p,\,\zeta)$, and the 
standard  polarization vectors, $e^\mu(\p,\ell)$, 
\beq
&& e^\mu(\p,\, \bot_1)={i\over \sqrt{2}}\Big(A^\mu(\p,\,3)- A^\mu(\p,\,1)
\Big)\,,\nonumber\\
&& e^\mu(\p,\, \bot_2)=-\,{1\over \sqrt{2}}
\Big(A^\mu(\p,\,3)+ A^\mu(\p,\,1)\Big)\,,\nonumber\\\nonumber\\
&& e^\mu(\p,\, \|)= i A^\mu(\p,\,2)\,,\nonumber\\
&& e^\mu(\p,\, 0)= i A^\mu(\p,\,4)\,,\label{e}
\eeq
where the symbols $\bot_1$ and $\bot_2$ represent two mutually orthogonal
polarizations perpendicular to $\p$, while the $\|$ labels polarizations
along the $\p$, and $0$ represents a time-like polarization.
The matrix of the momentum-space field strength is found to 
be,\footnote{As is apparent from Table 1,  $e^\mu(\p,\,0)$ is nothing 
but  $(1/\sqrt{m})\; p^\mu$. Also, of interest
is to note that the factors of, $i$, 
in Eqs. (\ref{e}), make $e^\mu(\p,\,\ell)$ real. } 
\beq
\F(\p,\,\ell)= e(\p,\,\ell) \;e(\p,\,0)^\dagger - e(\p,\,0) \;
e(\p,\,\ell)^\dagger\,.
\eeq                                                                 
where $\ell=\bot_1,\bot_2,\|,0$. This definition 
satisfies
\beq
\F(\p,\,\ell)\, \g\, e(\p,\,0) = m \;e (\p,\,\ell)\,,\quad\ell
=\bot_1,\bot_2,\|\,.
\eeq
The above equation is equivalent to the Proca equation in momentum space. 

The explicit expressions for $\epsilon^\mu(\p,\,\ell)$, and the 
corresponding $\F^{\mu\nu}(\p)$,
are
summarized in Table 1.


\def\cp{c_\phi}
\def\ct{c_\theta}
\def\sp{s_\phi}
\def\st{s_\theta}


\def\ex{ \sqrt{m}\;\left(
\begin{array}{c}
0\\
-\cp\ct\\
-\sp\ct\\
\st
\end{array}
\right)
}

\def\ey{\sqrt{m}\left(
\begin{array}{c}
0\\
 \sp\\
-\cp\\
0
\end{array}
\right)
}
\def\ez{\frac{1}{\sqrt{m}}
\left(
\begin{array}{c}
p\\
E\,\cp\;\st\\
E\,\sp\;\st\\
E\,\ct
\end{array}
\right)
}
\def\e0{\frac{1}{\sqrt{m}}
\left(
\begin{array}{c}
E\\
p\, \cp \st\\
p\, \sp \st\\
p\, \ct
\end{array}
\right)
}

\def\fx{
\left(
\begin{array}{cccc}
0 & E\, \cp \ct & E\, \ct \sp & - E\, \st \\
-E  \,\cp \ct & 0 & 0 & -p \,\cp \\
-E \,\ct \sp  & 0 & 0 & -p\, \sp\\
E \,\st & p \,\cp & p\, \sp &0
\end{array}
\right)
} 


\def\fy{
\left(
\begin{array}{cccc}
0 & -E\,\sp & E \,\cp & 0 \\
E\,\sp & 0 & p\,\st & p\,\ct\,\sp\\ 
-E \cp & -p\,\st & 0 & -p\,\cp\ct \\
0 & -p\,\ct\,\,\sp & p\, \cp\,\ct & 0
\end{array}
\right)
}

\def\fz{
\left(
\begin{array}{cccc}
0 & -m\,\cp\,\st & -m\sp\,\st & -m \ct \\
m \,\cp\,\st & 0 & 0 & 0\\
m\,\sp\,\st & 0 & 0 & 0\\
m\,\ct & 0 & 0 & 0
\end{array}
\right)
}


\def\f0{
\left(
\begin{array}{cccc}
0&0&0&0\\
0&0&0&0\\
0&0&0&0\\
0&0&0&0
\end{array}
\right)
} 


\def\fmunu{
\left(
\begin{array}{cccc}
0 & E_x (\p)& E_y(\p) & E_z(\p)\\
-E_x (\p)& 0 & -B_z(\p) & B_y(\p) \\
-E_y (\p)& B_z(\p) & 0 & -B_x(\p)\\
-E_z(\p) & -B_y(\p) & B_x(\p) & 0  
\end{array}
\right)
}

\begin{center}
\begin{table}
\begin{tabular}{|c|c|c|}\hline
$\ell$          & $e(\p\,,\ell)$  & $\F(\p\,,\ell) $   \\ 
\hline\hline
$\bot_1$ & $\ex$ & $\fx$  \\ 
\hline
$\bot_2$ & $\ey$ & $\fy$ \\ 
\hline		
$\|$     & $\ez$ & $\fz$ \\
\hline
$0$      & $\e0$ & $\f0$\\
\hline
\end{tabular}
\vspace{1.0cm}
\caption{Field strength tensor in momentum space for each of the polarization 
vectors. We have used the abbreviations:
$\cos(x)=c_x$, $\sin(x)=s_x$, $p=\vert\p\vert$.
The $  e^\mu(\p,\,\ell)$ are divergence-free for  $\ell=\bot_1,\bot_2,\|$,
while the contrary is true for $\ell=0$:
$p_\mu e^\mu(\p,\,\ell)=0$, for $\ell=\bot_1,\bot_2,\|$; while,
$p_\mu e^\mu(\p,\,\ell)= m^{3/2}$, for $\ell=0$.
}
\vspace{1.5cm}
\label{Table}
\end{table}
\end{center}

There is nothing in our formalism which requires to identify
$\F(\p\,,\ell) $ with massive electrodynamics. However,
doing so allows to make a few consistency tests and to gain 
a few useful insights.
Thus, identifying the $\F(\p\,,\ell) $ matrix as following general matrix,
\beq
\F^{\mu\nu}(\p) =\fmunu\,,
\eeq
it is apparent that the following consistency tests are satisfied:
\begin{enumerate}

\item
From Table 1, we immediately infer that all components of the
$\B(\p)$ fields are proportional to the magnitude of $\p$. Thus,
it verifies that without currents, i.e. with \p=\0, there are no
$\B$ components in  $\F(\p\,,\ell) $.

\item
In the massless limit, the longitudinal  $\F(\p\,,\ell) $, i.e.
 $\F(\p\,,\ell=\|) $,
identically vanishes.

\item
In the massless limit, setting  $\theta=\pi/2$, and $\phi=\pi/2$, 
yields the expected  $\F(\p\,,\ell) $  for an electromagnetic
wave propagating along the $y$-axis. 
The resulting $\F(\p\,,\bot_1) $ (and 
$\F(\p\,,\bot_2) $) contain 
 $\E$ and $\B$ fields which are respectively
along the $z$ ($x$)- and $x$ ($z$)- axes. Furthermore, 
they carry equal magnitudes.

\end{enumerate}

{\bf 4.} {\sc A conjecture and concluding remarks }

Having gathered together all the essential elements
we come to the final task of this essay.
We assert that the essential result, and the lesson to be learned from 
the renormalizable electroweak theory, is that  
massive gauge theories based upon Proca's framework are incomplete.
On the one side, the renormalizability of the electroweak theory demands that 
Proca framework be supplemented by the St\"uckelberg sector.
The latter,  in the usual framework, is 
introduced via a scalar field (or, fields).
On the other side, completeness in  $(1/2,1/2)$ 
requires it to also consist of two sectors.
At rest, one of them, the time-like polarization vector,
behaves as a scalar, while the remaining three degrees of freedom
constitute an ordinary spin-one vector.
In boosted inertial frames, the former ``scalar'' from the rest frame 
transforms into a state that is no longer of specified spin
and that does not transform according to the trivial representation
of the Lorentz group. It is that very state that produces the
St\"uckelberg term. 
It is worth emphasizing that
the time-like polarizations associated with 
$\A^\mu(\p,\,4)$, or, equivalently with $e^\mu(\p,0)$,  
do not contribute to  $\F^{\mu\nu}(\p\,,\ell)$. 
That is, the St\"uckelberg sector does not produce classical 
forces. Its importance lies  in its contribution to
quantum phases, to the propagator, and its role in renormalizability of
massive gauge theories. 

The relevant representation space for gravitational phenomena -- dictated
by the spacetime metric -- in being a direct product of two  
\hh representation spaces, in local inertial frames, carries counterparts
of all the above-indicated mathematical and physical elements.

Quantum theory of gravity should be expected to be a 
supersymmetric theory of fuzzy spacetime entities with a set of gravitational 
gauge bosons. The latter set contains, at the very least, 
massive gravitinos, and  nearly massless gravitons as 
dictated by observational data.

A detailed spacetime structure of massive gravitinos
can be found in Ref. \cite{ka}. There it is argued that
spin measurement on a massive gravitino shall yield not only
the expected spin component of spin $3/2$, but also an additional
set of components carrying spin $1/2$. These components, on the
basis of the lessons learned from the electroweak theory, 
should be expected to carry
significance for the renormalizability of the theory.

The other gravitational gauge boson, graviton, 
transforms as a massive $(1/2,1/2)\otimes(1/2,1/2)$ particle without
projecting out the lower spin components -- i.e., the St\"uckelberg 
counterparts. In keeping the lower spin components of the massive 
gravitational bosons, we produce a framework which differs
from the one to which the van Dam-Veltman considerations  
apply \cite{vanDam,mv}.  Such a field, in its rest frame, 
contains $16$ degrees of freedom (dof) distributed over 
a single spin-$2$ component ($5$ dof), 
three spin-$1$ sectors ($9$ dof),
and two spin-$0$ components ($2$ dof). 
Furthermore for CPT invariance one must also incorporate the
charge conjugated degrees of freedom. 
Thus, in a CPT covariant structure of the massive 
$(1/2,1/2)\otimes(1/2,1/2)$ contains 
a spin-$2$ component with $10$ dof, three  
spin-$1$ sectors with $18$ dof,
and two spin-$0$ components with $4$ dof.

Once the van Dam-Veltman observations no longer apply, one may
look at the graviton as an object described by a
massive $(1/2,1/2)\otimes(1/2,1/2)$ 
representation space.  The graviton, then, is not a single
spin object, but it carries in it the several spin components.
This is in exact parallel  of the gauge bosons of the electroweak theory
where the bosons cannot be seen as pure spin objects 
(except in their rest frame).

We thus make the following conjecture:

\begin{quote}
The renormalizability of the quantum theory of gravity would require 
that the gravitational bosons, gravitinos and graviton, 
be  treated as multi-spin particles in the sense defined above.  
\end{quote}

This conjecture has some remarkable additional consequences.
To see this, first note that 
despite the fact that
the full \hh rep\-resen\-tation space is spanned by particles
which do not carry a definite spin as encoded in the
spin content of the $\A^\mu(x)$, the associated 
curvature, as encoded in the $\F^{\mu\nu}(x)$, the field strength tensor,
is a pure spin one object. 
Similarly, despite the fact that 
the massive graviton,
transforming as $(1/2,1/2)\otimes(1/2,1/2)$ rep\-resenta\-tion space,
carries several spin components, the associated field
strength tensor $\R^{\mu\nu\lambda\sigma}(x)$ -- 
i.e. the Riemannian curvature tensor --
is a pure spin-$2$ object. 

The incomplete treatments of the $\A^\mu(x)$, as well 
as $\psi^\mu(x)$ and $g^{\mu\nu}(x)$,  either ignore, or project out, 
lower spin components. However, the latter  
are natural inhabitants of these spaces. Yet, this circumstance does 
not affect the induced curvatures, $\F(x)$ and $\R(x)$. 
As such, as far as measured forces are
concerned the noted incompleteness carries  no significance.  
The heuristic treatments of   $(1/2,1/2)$, as well 
as $(1/2,1/2)\otimes(1/2,1/2)$ yield correct
electroweak and 
gravitational {\em forces\/}.
The lower spin
components enter entirely 
at the quantum level -- without inducing
forces -- by their contribution residing in certain phases,
and via their contributions to the propagator that makes 
the theories renormalizable.
The latter phases can be studied via
neutrino oscillations, or a set of orthogonal  states in linear 
superposition of different energy/mass eigenstates
 \cite{grf1996,prd1998,kk,jw,grf1997,Ryder}.\footnote{For an 
early work on gravitationally induced phases in neutron interferometry
\cite{cow}, see Ref. \cite{ja}.
With notable exceptions of Ref. \cite{ls,mg}, most early works
on gravitationally induced phases were devoted to single mass
eigenstates. The above-quoted references address themselves to 
states in linear superposition of different mass eigenstates. It
allows to probe certain additional, and non-trivial, gravitationally induced
phases. }

Thus a quantum theory of gravity lives in a non-commutative spacetime,
and in fact is its theory. Furthermore, even when the non-commutative 
structure of spacetime can be neglected, the conceptual framework 
requires attention not only to forces that are acted upon test states, 
but one must also pay due attention to existence of certain non-trivial 
gravitationally-induced quantum phases. 
Under certain circumstances the former may be zero, while the latter
may be non-vanishing and observable.
To run the point home, first recall that local density fluctuations 
in the cosmological context can, and do, create regions characterizable 
by a set of dimensionless gravitational potentials. The latter have a 
characteristic dimensionless value of about $\vert\Phi_0\vert
\sim10^{-5}$, and are, in general, several order of magnitude  
larger than those arising from stars and planets that may inhabit these
regions. In our solar system the (magnitude of)
dimensionless lunar 
gravitational potential is $3.14\times 10^{-11}$, for Earth it
is $6.95\times 10^{-10}$, while for Sun it is $2.12\times 10^{-6}$.
Next, note that while the planetary and lunar 
orbits are determined by gradients in these latter potentials; the 
$\Phi_0$, to a good approximation, 
essentially has the affect of red-shifting the orbital periods.
However, for quantum systems embedded in quantum/classical 
gravitational fields one may entertain a violation of the equivalence
principle (VEP) with observable phenomenological 
effects \cite{vep_obs,vep_obs2,vep_obs3}.
In such scenarios,  the essentially force-free 
 $\vert\Phi_0\vert
\sim10^{-5}$ amplifies the terrestrial observability
of VEP
 by about five orders
of  magnitude.

In reference to Eq. (\ref{mwpd}), one need not await to
reach Planck energies in terrestrial accelerators.
Quantum states carrying Planck mass can be easily 
created and studied in laboratory using superconducting quantum interference
devices (SQUID). 
In these devices the mass carried by the superconducting quantum state
is,
\beq
M_{SQUID}\sim f(T)\, m_c\, N_A\,,
\eeq
where $m_c$ is the mass of a Cooper pair, $N_A$ is the Avogadro number,
and $f(T)$ is the fraction of the electrons that are in a superconducting 
state. Since all the Copper pairs are part of a single superconducting state,
and $f(T)$ can reach close to unity at temperatures, $T$, sufficiently below
the critical temperature, $M_{SQUID}$ becomes of the order of Planck mass.
This fact has apparently escaped attention. But it may carry significance 
for theorists as well as experimentalists to probe the interface of the
gravitational and quantum realms in the emerging field of 
experimentally-driven quantum gravity
phenomenology.

\vspace{1.5 cm}
{\em This work was supported by Consejo Nacional de Ciencia y Tecnolog\'ia
(CONACyT) under grant number 32067-E.\/}


\begin{thebibliography}{000}

\bibitem{as}
R.\ J.\ Adler, D.\ I.\ Santiago,
{\em On gravity and uncertainty principle,\/}
Mod. Phys. Lett. A {\bf 14}, 1371 (1999).

\bibitem{grf1994}
D.\ V.\ Ahluwalia,
{\em Quantum measurement, gravitation, and locality,}
Phys. Lett. B {\bf 339}, 301 (1994).

\bibitem{akempf}
A.\ Kempf, G.\ Mangano, R.\ B.\ Mann,
{\em Hilbert space representation of the minimal uncertainty relation,\/}
Phys. Rev. D {\bf 52}, 1108 (1995).  

\bibitem{pla2000}
D.\ V.\ Ahluwalia,
{\em Wave-Particle duality at the Planck scale: Freezing of neutrino 
oscillations,\/} 
Phys. Lett. A {\bf 275}, 31 (2000). 

\bibitem{jm}
J.\ Madore, 
{\em Gravity on fuzzy space-time,\/}
Los Alamos Archive preprint
gr-qc/9709002.

\bibitem{sh}
S.\ de Haro,
{\em Non-commutative black hole algebra and string theory from gravity,\/}
Class. Quant. Grav. {\bf 15}, 519 (1998).

\bibitem{gac}
G.\ Amelino-Camelia,
{\em Classicality, matter-antimatter asymmetry, and quantum\\-gravity
deformed uncertainty relations,\/}
Mod. Phys. Lett. A {\bf 12}, 1387 (1997).

\bibitem{mm}
M.\ Maggiore,
{\em A generalized uncertainty principle in quantum gravity,\/} 
Phys. Lett. B {\bf 304}, 65 (1993).

\bibitem{ns}
N.\ Sasakura,
{\em An uncertainty relation of space-time,\/}
Prog. Theor. Phys. {\bf 102}, 169 (1999).

\bibitem{fs}
F.\ Scardigli,
{\em Generalized uncertainty principle in quantum gravity from micro-black hole
gedanken experiment,\/}
Phys. Lett. B {\bf 452}, 39 (1999).

\bibitem{gacn}
G.\ Amelino-Camelia,
{\em An interferometric gravitational wave detector as a quantum-gravity
apparatus,\/}
Nature {\bf 398}, 216 (1999).

\bibitem{dvan}
D.\ V.\ Ahluwalia,
{\em Quantum gravity: Testing time for theories,\/}
Nature {\bf 398}, 199 (1999).



\bibitem{hk}
H.\ V.\ Klapdor-Kleingrothaus,
A.\ Dietz, H.\ L.\ Harney, I.\ V.\ Krivosheina,
{\em Evidence for neutrinoless double beta decay,\/}
Mod. Phys. Lett. A {\bf 16}, 2409 (2001).

\bibitem{ew}
E.\ Witten,
{\em High-energy physics: The mass question,\/} 
Nature {\bf 415}, 969 (2002).

\bibitem{ak}
D.\ V.\ Ahluwalia, M.\ Kirchbach,
{\em Particle-antiparticle metamorphosis of massive Majorana neutrinos and
gauginos,\/} Phys. Lett. B (submitted, 2002).


\bibitem{Veltman}
M.\ J.\ G.\ Veltman,
{\em From weak interactions to gravitation,\/}
Int. J. Mod. Phys. A {\bf 15}, 4557 (2000).

\bibitem{acc}
A.\ G.\ Riess {\em et al.\/}
{\em The farthest known supernova: Support for an accelerating
universe and a glimpse of the epoch of deceleration,\/}
Astrophys. J. {\bf 560}, 49 (2001).

\bibitem{tm}
T.\ Matos, F.\ S.\ Guzman,
{\em On the spacetime of a galaxy,\/}
Class. Quant. Grav. {\em 18}, 5055 (2001).

\bibitem{ds}
 U.\ Nucamendi, M.\ Salgado, D.\ Sudarsky
{\em An alternative approach to the galactic dark matter problem,\/}
Phys. Rev. D {\bf 63},  125016 (2001).

\bibitem{nd1}
N.\ Dadhich, N.\ Banerjee,
{\em Global monopoles and scalar fields as the electrogravity dual 
of Schwarzschild space-time,\/}
Mod. Phys. Lett. A {\bf 16}, 1193 (2001).

\bibitem{nd2}
N.\ Dadhich, 
{\em On electrogravity duality,\/}
Mod. Phys. Lett. A {\bf 14}, 337 (1999).

\bibitem{fs2}
L.\ S.\ Finn, P.\ J.\ Sutton,
{\em Bounding the mass of the graviton using pulsar observations,\/}
Phys. Rev. D {\bf 65}, 044022 (2002).


\bibitem{ab}
Y.\ Aharonov, D.\ Bohm,
{\em Significance of electromagnetic potentials in the quantum theory,\/}
Phys. Rev. {\bf 115}, 485 (1959).


\bibitem{ka}
M.\ Kirchbach, D.\ V.\ Ahluwalia,
{\em Spacetime structure of massive gravitino,\/}
Phys. Lett. B {\bf 529}, 124 (2002).



\bibitem{bww}
D.\ V.\ Ahluwalia, M.\ B.\ Johnson, T.\ Goldman,
{\em A Bargmann-Wightman-Wigner type quantum field theory,\/}
Phys. Lett. B {\bf 316}, 102 (1993).


\bibitem{vanDam}
H.\ van Dam, M.\ Veltman,
{\em Massive and massless Yang-Mills and gravitational fields,\/} 
Nucl. Phys. B {\bf 22}, 397 (1970). 

\bibitem{mv}
M.\ Visser,
{\em Mass for the Graviton,\/}
Gen. Rel. Grav. {\bf 30}, 1717 (1998).


\bibitem{grf1996}
D.\ V.\ Ahluwalia, C.\ Burgard,
{\em Gravitationally induced neutrino-oscillation phases,\/}
Gen. Rel. and Grav. 28, 1161 (1996); Errata: 29,681 (1997).


\bibitem{prd1998}
 D.\ V.\ Ahluwalia, C.\ Burgard,
{\em Interplay of gravitation and linear superposition 
of different mass eigenstates,\/} 
Phys. Rev. D {\bf 57}, 4724 (1998). 


\bibitem{kk}
K.\ Konno, K.\ Kasai,
{\em General relativistic effects in gravity and quantum mechanics: a
case of ultra-relativistic, spin 1/2 particles,\/}
Prog. Theor. Phys. {\bf 100}, 1145 (1998).

\bibitem{jw}
J.\ Wudka,
{\em Mass dependence of the gravitationally induced wave-function phases,\/}
Phys. Rev. D {\bf 64}, 065009 (2001).

\bibitem{grf1997}
D.\ V.\ Ahluwalia,
{\em On a new non-geometric element in gravity,\/} 
       Gen. Rel. and Grav. {\bf 29}, 1491 (1997). 

\bibitem{Ryder}
M.\ Adak, T.\ Dereli, L.\ H.\ Ryder,
{\em Neutrino oscillations induced by space-time torsion,\/}
Class. Quant. Grav. {\bf 18}, 1503 (2001).

\bibitem{cow}
R.\ Colella, A.\ W.\ Overhauser, S.\ A.\ Werner, 
{\em Observation of gravitationally induced quantum interference.\/}
Phys. Rev. Lett. {\bf 34}, 1472 (1975).

\bibitem{ja}
J.\ Anandan, {\em Gravitational and rotational effects in
quantum measurement,\/} Phys. Rev. D {\bf 15}, 1448 (1997).


\bibitem{ls}
L.\ Stodolsky,
{\em Matter and light wave interferometry in gravitational fields,\/}
Gen. Rel. Grav. {\bf 11}, 391 (1979).

\bibitem{mg}
M.\ L.\ Good,
{\em $K_2^0$ and the equivalence principle,\/}
Phys. Rev. {\bf 121}, 311 (1961).


\bibitem{vep_obs}
R.\ Horvat,
{\em Pulsar velocities due to a violation of the 
equivalence principle by neutrinos,\/}
Mod.Phys.Lett. A {\bf 13}, 2379 (1998).

\bibitem{vep_obs2}
M.\ Barkovich, H.\ Casini, J.\ C.\ D'Olivo, R.\ Montemayor,
{\em Pulsar motions from neutrino oscillations induced 
by a violation of the equivalence principle,\/}
Phys. Lett. B {\bf 506,} 20 (2001).

\bibitem{vep_obs3}
G.\ Lambiase,
{\em Neutrino oscillations induced by gravitational recoil effects,\/}
Los Alamos Archive preprint gr-qc/0107066. 



\end{thebibliography}
\end{document}